\documentclass[11pt,a4paper]{article}
\usepackage{times,latexsym}
\usepackage{url}
\usepackage[T1]{fontenc}

\usepackage[acceptedWithA]{tacl2021v1}

\usepackage{xspace,mfirstuc,tabulary}

\usepackage{multirow,array,adjustbox}
\usepackage{caption}
\usepackage[table]{xcolor}
\usepackage{hyperref}
\usepackage{algorithm}
\usepackage{algpseudocode}
\usepackage{pgfplots}
\usepackage{xcolor}
\usepackage{colortbl}
\usepackage{amsmath} 
\usepackage{booktabs}
\usepackage{amssymb} 
\usepackage{dsfont}  

\definecolor{first}{RGB}{191, 225, 201} 
\definecolor{second}{RGB}{227, 237, 185} 
\definecolor{third}{RGB}{254, 250, 194} 
\definecolor{stab}{RGB}{232, 241, 252} 

\newif\iftaclinstructions
\taclinstructionsfalse 
\iftaclinstructions
\renewcommand{\confidential}{}
\renewcommand{\anonsubtext}{(No author info supplied here, for consistency with
TACL-submission anonymization requirements)}
\newcommand{\instr}
\fi

\iftaclpubformat 

\else

\fi

\title{Sci-Mind: Cognitively-Inspired Adversarial Debate for Autonomous Mathematical Modeling}

\author{
  Junhao Jia$^{\diamond, \ddagger}$, 
  Huangwei Chen$^{\diamond, \ddagger}$, 
  Ruiying Sun$^\dagger$, 
  Yanhui Song$^\ddagger$, 
  Haishuai Wang$^{\diamond,}$\Thanks{Corresponding authors.}, 
  Jiajun Bu$^\diamond$, 
  Lei Wu$^{\diamond,*}$ 
  \\[1ex] 
  $^\diamond$Zhejiang Key Laboratory of Accessible Perception and Intelligent Systems, \\
  College of Computer Science and Technology, Zhejiang University, China \\[1ex] 
  \texttt{junhaojia530@gmail.com}, \texttt{\{haishuai.wang, shenhai1895\}@zju.edu.cn}
  \\[1ex] 
  $^\dagger$Hangzhou Normal University, China \quad\quad $^\ddagger$Hangzhou Dianzi University, China
}

\date{}

\begin{document}
\maketitle
\begin{abstract}
Real-world mathematical modeling is inherently an experiential and collaborative endeavor. Domain experts rarely solve complex problems from scratch; instead, they draw upon analogies from historical cases and subject their hypotheses to rigorous peer scrutiny. However, autonomous agents powered by Large Language Models predominantly rely on isolated reasoning paradigms, frequently generating plausible but fundamentally flawed models due to a lack of domain grounding and adversarial verification. To address these limitations, we propose Sci-Mind, a novel framework that mirrors the human scientific discovery process. Sci-Mind integrates Experiential Memory Recall to retrieve executable code snippets and modeling paradigm descriptors, grounding abstract reasoning in historical solutions. Subsequently, it employs an Adversarial Cognitive Dialectic where a Theorist optimizing mathematical coherence and a Pragmatist enforcing data feasibility debate through competing objectives to prune elegant but infeasible formulations. A Self-Validating Execution Strategy further ensures blueprint consistency through formal predicates before code generation, achieving fully autonomous execution. Extensive experiments on the MM-Bench and EngiBench demonstrate that Sci-Mind significantly outperforms leading autonomous agents in both modeling rigorousness and code executability.
\end{abstract}

\section{Introduction}

Mathematical modeling serves as the fundamental bridge translating complex physical phenomena into formal analytical systems~\cite{hao2024pinnacle}. Despite its wide applications in operations research~\cite{bengio2021machine}, physical simulations~\cite{karniadakis2021physics}, and biomedical network analysis~\cite{jia2025brain,jia2025rtgmff}, autonomous tracking of open-ended scientific tasks relying solely on isolated reasoning often fails under complex real-world constraints~\cite{romera2024mathematical}. In contrast, human scientific methodology has the advantage of experiential and collaborative frameworks, providing rigorous verification and historical grounding~\cite{popper2005logic}. To integrate these advantages, augmenting Large Language Models (LLMs) with scientific reasoning paradigms is crucial for robust performance.

\begin{figure}[t]
\centering
\includegraphics[width=\columnwidth]{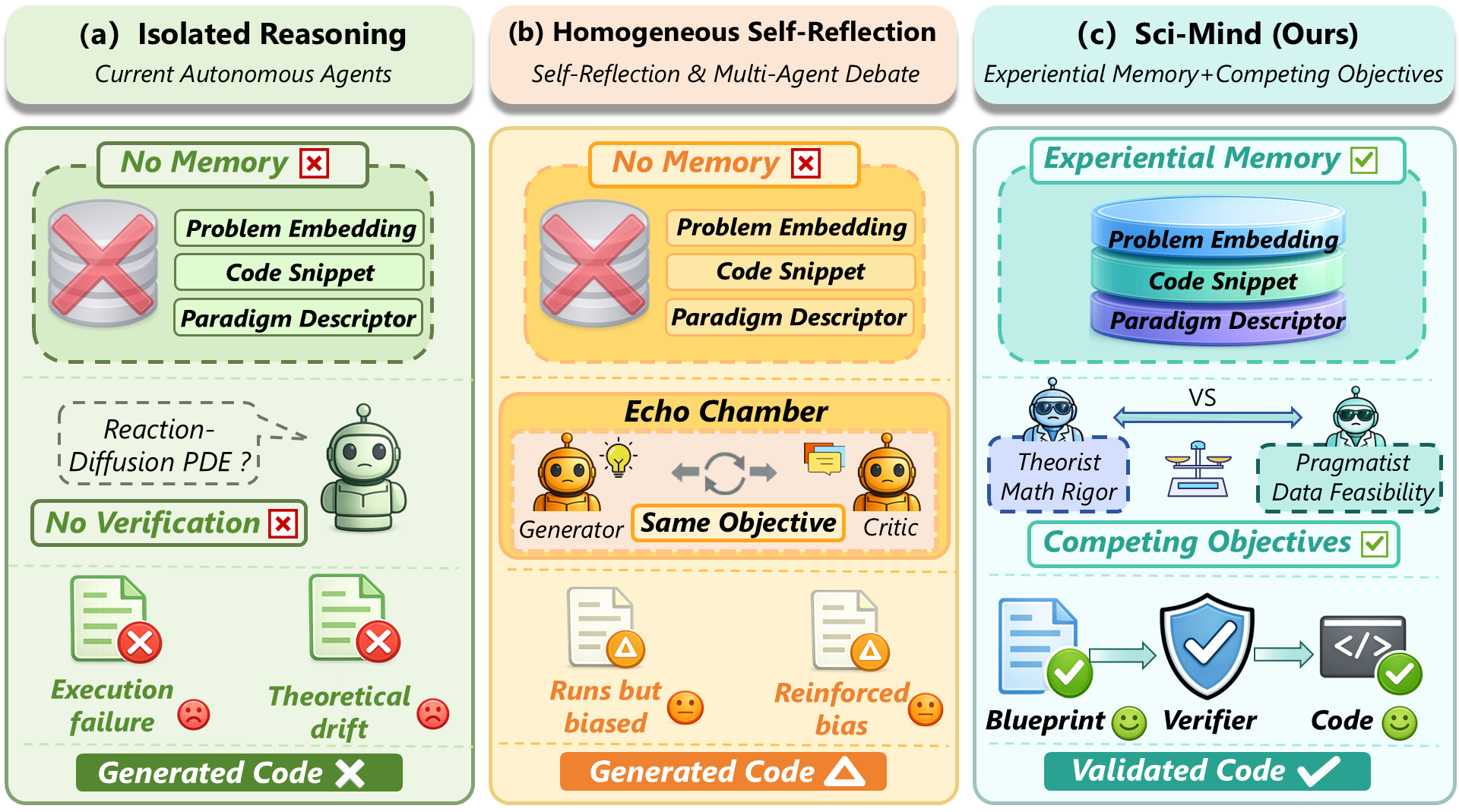}
\caption{Comparison of three paradigms for LLM-based mathematical modeling.} 
\label{fig1}
\end{figure}

However, current autonomous agents model the problem using zero-shot inference from the initial prompt, which often leads to severe theoretical drift or execution failures~\cite{jimenez2024swebench}. This limitation stems from two main factors. First, a single isolated reasoning process offers limited structural information about the target problem, failing to capture complex modeling paradigms across different scientific domains~\cite{lu2023mathvista}. Second, the inherent ambiguity of real-world data often causes agents to overemphasize elegant but unvalidated theoretical features, resulting in practical infeasibility~\cite{yin2024mumath}. Moreover, directly translating abstract math into code without structural priors is highly brittle in practical scenarios~\cite{lai2023ds1000}, consequently hindering its widespread deployment. As illustrated in Fig.~\ref{fig1}(a), these agents operate without experiential memory or adversarial verification, causing both execution failures and theoretical drift in the generated code.

On the other hand, recent advances in LLM reasoning have focused on integrating self-reflection mechanisms~\cite{pan2024automatically}. In previous works, actor and critic models are first prompted independently, then fused through iterative feedback loops operating at the semantic level~\cite{reflexion}. However, a significant portion of this self-reflection is redundant, forming an echo chamber that amplifies generative biases without exposing logical blind spots~\cite{valmeekam2023planning}. As shown in Fig.~\ref{fig1}(b), when the generator and critic share the same objective, the resulting code may execute but carries reinforced biases that remain undetected. Furthermore, heterogeneous domain gaps between abstract theoretical logic and concrete programming syntax hinder the establishment of effective code execution~\cite{zan2023large}. This raises a basic question: \textit{How can we achieve deep theoretical rigor while escaping echo-chamber redundancies and bridging the gap to executable code?}

To address these challenges, we introduce cognitive bionics as a high-level representation to enhance mathematical modeling in LLMs. Human-like scientific discovery provides a more grounded understanding of problems than isolated reasoning, addressing structural limitations in abstract generation~\cite{langley1987scientific}. Experiential memory representation can offer clear semantic priors, including logic paradigms and executable demonstrations, enabling effective theory-implementation separation~\cite{kolodner1992introduction}. Recent advances in Retrieval-Augmented Generation (RAG) have demonstrated the potential of external knowledge for robust reasoning~\cite{rag}. However, these methods face the challenge of semantic misalignment, as they predominantly fetch unstructured text that fails to map onto formal executable structures~\cite{barnett2024seven}. To bridge this gap, we extend current retrieval methods by introducing executable code-level insights and a non-cooperative game mechanism.

Building upon the cognitive bionic paradigm, we propose a novel mathematical modeling framework. As depicted in Fig.~\ref{fig1}(c), our framework consists of three key components: Experiential Memory Recall (EMR), Adversarial Cognitive Dialectic (ACD), and a Self-Validating Execution Strategy (SVE). The EMR retrieves executable code snippets and modeling paradigm descriptors from a structured knowledge base, grounding abstract reasoning in historical solutions. The ACD employs functionally asymmetric personas, a Theorist and a Pragmatist, who optimize explicitly competing objectives to dynamically verify both theoretical validity and data feasibility. The SVE implements an automated blueprint-verifier-code pipeline with structured JSON blueprints and formal consistency predicates to bridge validated hypotheses and code execution.

With the above components, our Sci-Mind enables comprehensive scientific reasoning, leveraging both historical evidence and adversarial context to maintain theoretical rigor under complex scenarios. Extensive experiments on the MM-Bench~\cite{mm_agent} and EngiBench~\cite{engibench} datasets demonstrate the effectiveness of our method. In summary, our main contributions are as follows:

(1) We propose Sci-Mind, a cognitively-inspired framework that retrieves executable structural priors from a multi-tier knowledge base via Experiential Memory Recall, effectively bridging the abstraction-implementation gap in autonomous mathematical modeling.

(2) We introduce the Adversarial Cognitive Dialectic, in which a Theorist and a Pragmatist debate through explicitly competing objectives, and demonstrate that this objective asymmetry is the critical factor for escaping theoretical local optima and echo-chamber redundancies.

(3) Extensive experiments across two benchmarks across multiple LLM backbones confirm that Sci-Mind consistently outperforms existing autonomous agents in both modeling rigorousness and code executability, with gains that are orthogonal to backbone capabilities.

\section{Related Work}

\subsection{Scientific Reasoning Agents}
LLM-powered autonomous agents have become a prominent paradigm for complex analytical tasks~\cite{wang2024survey,react}, progressing from Zero-shot CoT prompting~\cite{cot,kojima2022large} to specialized workflows such as DS-Agent~\cite{ds_agent} for data science and MM-Agent~\cite{mm_agent} for mathematical modeling, as well as domain-specific scientific agents like Coscientist~\cite{boiko2023coscientist} and ChemCrow~\cite{bran2024chemcrow}. While integrating RAG~\cite{rag} helps mitigate hallucinations~\cite{gao2024rag_survey}, existing retrieval methods predominantly fetch unstructured text that fails to bridge abstract formulations and executable code. Sci-Mind addresses this gap through Experiential Memory Recall, which retrieves executable code snippets alongside modeling paradigm descriptors, providing both procedural and conceptual priors.

\subsection{Self-Reflection and Multi-Agent Debate}
Self-reflection methods such as Reflexion~\cite{reflexion} and Self-Refine~\cite{selfrefine} enable iterative critique within a single model, while multi-agent frameworks, including CAMEL~\cite{camel}, MetaGPT~\cite{metagpt}, FetalAgents~\cite{hu2026fetalagents}, and AutoGen~\cite{autogen}, extend this idea through collaborative role-based interaction. Multi-agent debate~\cite{debate,liang2024encouraging,chan2024chateval} further improves robustness via multi-round discussion. However, when all agents share the same implicit objective, the interaction tends to degenerate into an echo chamber that reinforces biases rather than exposing fundamental flaws~\cite{huang2024large}. This is especially problematic in scientific modeling, where theoretical validity and data feasibility are structurally distinct evaluation dimensions. Sci-Mind addresses this through the Adversarial Cognitive Dialectic, in which a Theorist and a Pragmatist debate with explicitly competing objectives.

\subsection{Code Generation and Execution}
Code generation research spans evaluation benchmarks~\cite{codegen}, runtime verification~\cite{ldb}, self-improving code libraries~\cite{voyager}, hierarchical task decomposition~\cite{data_interpreter}, and sandboxed agentic execution~\cite{mm_agent,engibench}. Despite these advances, direct logic-to-code translation remains brittle when modeling and implementation decisions are entangled in a single generation pass. Sci-Mind decouples these concerns through a Self-Validating Execution Strategy that interposes a structured JSON blueprint and an automated Verifier with formal consistency predicates before code generation.

\begin{figure*}
    \centering
    \includegraphics[width=0.95\linewidth]{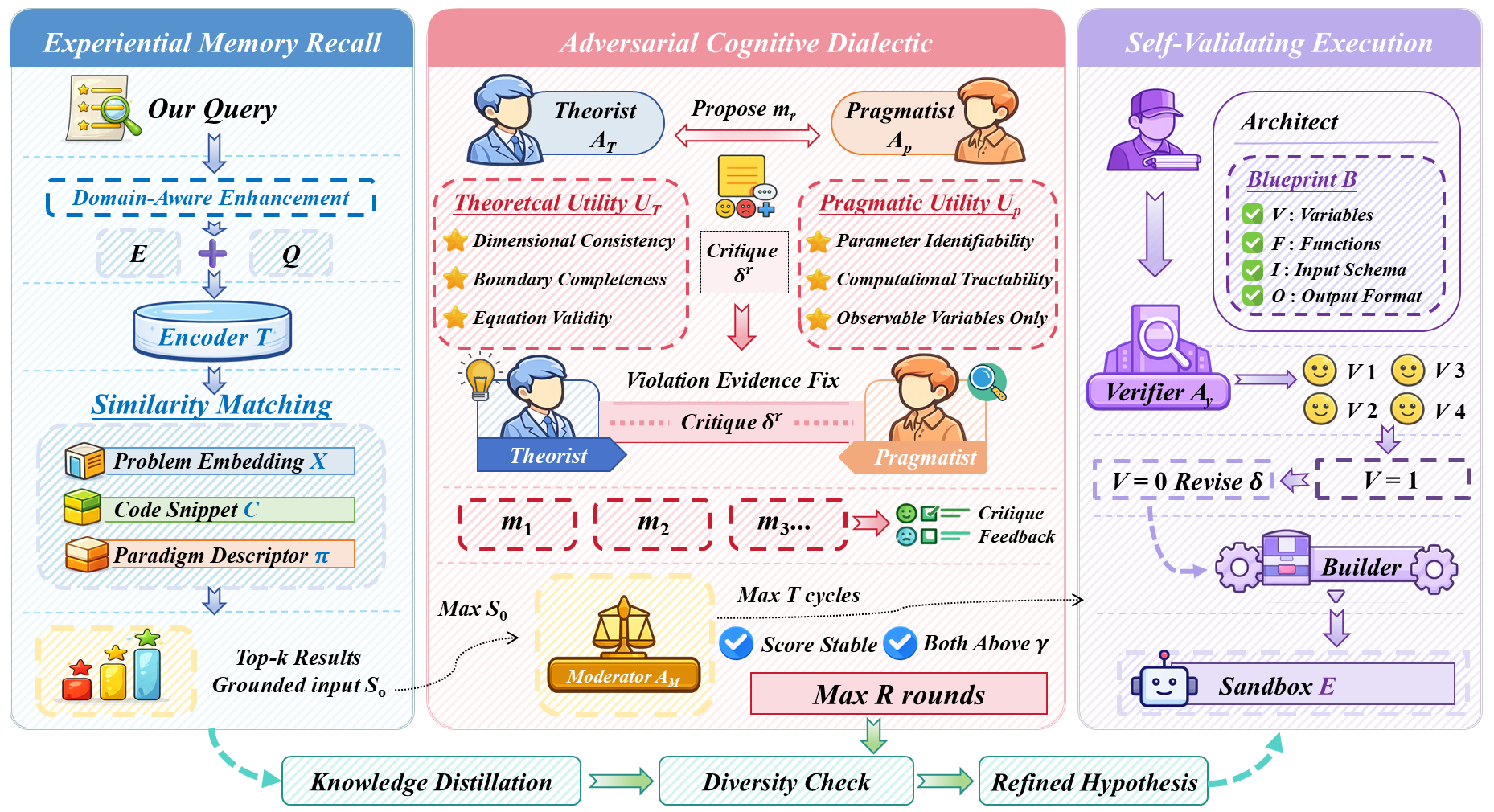}
\caption{The Overall Architecture of the Sci-Mind Framework.} \label{fig2}
\end{figure*}

\section{Methodology}

In this paper, we propose Sci-Mind, a cognitively-inspired framework for autonomous mathematical modeling, as shown in Fig~\ref{fig2}.
 
Formally, given a user query $\mathbf{Q} \in \mathbb{R}^{N \times d}$ where $N$ is the token length and $d$ is the hidden dimension, the overall pipeline is formulated as a sequential composition of three mappings:
\begin{equation}
  \mathbf{C}^{*} = \mathcal{S} \circ \mathcal{A} \circ \mathcal{R}(\mathbf{Q}, \mathcal{K})
\end{equation}
where $\mathcal{R}$ denotes the EMR retrieval operator, $\mathcal{A}$ represents the ACD refinement process, and $\mathcal{S}$ is the SVE code synthesis module. The knowledge base $\mathcal{K}$ is maintained as a dynamic repository of structured triplets, and the output $\mathbf{C}^{*}$ is an executable code artifact validated within a sandboxed environment $\mathcal{E}$. Details of each component are described as follows.

\subsection{Experiential Memory Recall (EMR)}

A critical bottleneck in autonomous modeling is the abstraction-implementation gap: an agent may possess sufficient declarative knowledge to formulate a model but lack the procedural knowledge to implement it as executable code. Existing retrieval-augmented methods~\cite{karpukhin2020dense} predominantly fetch unstructured textual definitions, which provide semantic context but fail to supply the syntactic scaffolding necessary for code generation. To bridge this gap, EMR retrieves structured historical solutions as executable priors.

\subsubsection{Multi-Tier Knowledge Representation}

We organize the knowledge base $\mathcal{K}$ as a set of structured triplets $\{k_i\}_{i=1}^{|\mathcal{K}|}$, where each entry $k_i = (\mathbf{x}_i, \mathbf{c}_i, \pi_i)$ consists of three components: the problem semantic embedding $\mathbf{x}_i \in \mathbb{R}^{d}$, a verified executable code snippet $\mathbf{c}_i$, and a modeling paradigm descriptor $\pi_i$ that encodes the abstract solution pattern. This tri-level representation ensures that retrieved knowledge simultaneously provides semantic context, implementation templates, and high-level structural guidance.

\subsubsection{Domain-Aware Retrieval}

To mitigate semantic misalignment between natural language problem descriptions and formal mathematical structures, we construct a domain-aware enhanced query. Specifically, we prepend a retrieval instruction prefix $\mathbf{E}$ to the input query, forming the augmented representation:
\begin{equation}
    \mathbf{H} = [\mathbf{E} \,;\, \mathbf{Q}]
\end{equation}
where $[\,;\,]$ denotes sequence concatenation and $\mathbf{E}$ is instantiated as ``\textit{Retrieve the formal mathematical structure for} $[\texttt{domain}]$'', with $[\texttt{domain}]$ dynamically populated from the query context.

The augmented query is encoded by a dense encoder $\mathcal{T}$ to produce a query vector $\hat{\mathbf{h}} \in \mathbb{R}^{d}$:
\begin{equation}
    \hat{\mathbf{h}} = \text{MeanPool}(\mathcal{T}(\mathbf{H}))
\end{equation}

We then compute the cross-modal relevance between $\hat{\mathbf{h}}$ and each knowledge entry. The relevance score for entry $k_j$ is defined as:
\begin{equation}
    s(\mathbf{Q}, k_j) = \frac{\hat{\mathbf{h}}^\top \phi(\mathbf{x}_j)}{\|\hat{\mathbf{h}}\|_2 \cdot \|\phi(\mathbf{x}_j)\|_2}
\end{equation}
where $\phi(\cdot)$ denotes the knowledge encoder, which shares parameters with $\mathcal{T}$ but operates on the stored problem embeddings.

The top-$k$ entries with the highest relevance scores are selected to form the structural prior set:
\begin{equation}
    \mathcal{C}_{top\text{-}k} = \underset{\mathcal{C} \subset \mathcal{K},\, |\mathcal{C}|=k}{\arg\max} \sum_{k_j \in \mathcal{C}} s(\mathbf{Q}, k_j)
\end{equation}

The retrieved code snippets and paradigm descriptors are concatenated with the original query to form the structurally-grounded input $\mathbf{S}^0$:
\begin{equation}
    \mathbf{S}^0 = [\mathbf{Q} \,;\, \bigoplus_{k_j \in \mathcal{C}_{top\text{-}k}} (\mathbf{c}_j \oplus \pi_j)]
\end{equation}
where $\oplus$ denotes token-level concatenation. This $\mathbf{S}^0$ serves as the grounded initialization for the subsequent adversarial refinement stage.

\subsection{Adversarial Cognitive Dialectic (ACD)}

Standard self-reflection methods prompt a single model to iteratively critique its own output. Recent studies~\cite{huang2024large} have shown that such homogeneous reflection frequently degenerates into an echo chamber that reinforces generative biases without exposing fundamental flaws. We argue that the root cause is objective homogeneity: both the generator and the critic optimize toward the same implicit objective. To address this, ACD decomposes the verification task into two explicitly competing objectives evaluated by functionally asymmetric agents, drawing on the principle that structured disagreement produces more robust outcomes than unchallenged consensus.

\subsubsection{Dual-Objective Formulation}

We define two evaluation functions that operate on different modalities of the modeling hypothesis. Given a candidate model hypothesis $\mathbf{m}$ and the observable dataset constraints $\mathcal{D}_{obs}$, we define:

\textbf{Theoretical Utility} $U_T(\mathbf{m})$ evaluates the mathematical coherence of the hypothesis, quantified as a weighted combination of interpretable criteria scored by the Theorist agent $\mathcal{A}_T$:
\begin{equation}
    U_T(\mathbf{m}) = \sum_{l=1}^{L} w_l^{(T)} \cdot \sigma_l^{(T)}(\mathbf{m})
\end{equation}
where $\sigma_l^{(T)}(\mathbf{m}) \in [0, 1]$ denotes the $l$-th evaluation criterion, $w_l^{(T)}$ is the corresponding importance weight with $\sum_l w_l^{(T)} = 1$, and $L$ is the number of criteria. Each $\sigma_l^{(T)}$ is implemented as a structured rubric prompt that returns a normalized score.

\textbf{Pragmatic Utility} $U_P(\mathbf{m} \mid \mathcal{D}_{obs})$ evaluates the feasibility of the hypothesis against concrete data constraints, scored by the Pragmatist agent $\mathcal{A}_P$:
\begin{equation}
    U_P(\mathbf{m} \mid \mathcal{D}_{obs}) = \sum_{l=1}^{L'} w_l^{(P)} \cdot \sigma_l^{(P)}(\mathbf{m}, \mathcal{D}_{obs})
\end{equation}
where $\sigma_l^{(P)}(\mathbf{m}, \mathcal{D}_{obs}) \in [0, 1]$ denotes pragmatic criteria, $w_l^{(P)}$ is the corresponding weight with $\sum_l w_l^{(P)} = 1$, and $L'$ is the number of pragmatic criteria.

The key design principle is that $U_T$ and $U_P$ are structurally non-aligned: maximizing $U_T$ alone tends to produce mathematically elegant but over-parameterized models, while maximizing $U_P$ alone leads to oversimplified models that lack generalization. The productive tension between these objectives drives the system toward balanced solutions.

\subsubsection{Structured Adversarial Interaction}

At debate round $r$, the interaction proceeds as a two-player sequential game:

\textbf{Step 1: Proposal.} The Theorist proposes or refines a model hypothesis conditioned on the structural prior and the accumulated critique history:
\begin{equation}
    \mathbf{m}^{r} = \mathcal{A}_T\left(\left[\mathbf{S}^0 \,;\, \boldsymbol{\delta}^{r-1}\right]\right)
\end{equation}
where $\boldsymbol{\delta}^{0} = \varnothing$ for the initial round.

\textbf{Step 2: Critique.} The Pragmatist evaluates $\mathbf{m}^r$ against $\mathcal{D}_{obs}$ and produces a structured critique identifying specific feasibility violations:
\begin{equation}
    \boldsymbol{\delta}^{r} = \mathcal{A}_P\left(\left[\mathbf{m}^r \,;\, \mathcal{D}_{obs}\right]\right)
\end{equation}

Each critique $\boldsymbol{\delta}^{r}$ is constrained to a structured format that specifies: (i) the violated pragmatic criterion, (ii) the conflicting evidence from $\mathcal{D}_{obs}$, and (iii) a suggested remediation direction. This structured format prevents vague or redundant feedback.

\textbf{Step 3: Moderation.} A Moderator agent $\mathcal{A}_M$ evaluates both utilities at each round and computes a joint consensus score:
\begin{equation}
    \Gamma^r = \lambda \cdot U_T(\mathbf{m}^r) + (1 - \lambda) \cdot U_P(\mathbf{m}^r \mid \mathcal{D}_{obs})
\end{equation}
where $\lambda \in (0,1)$ is a balancing coefficient that controls the trade-off between theoretical elegance and practical feasibility. The debate terminates when the following convergence condition is satisfied:
\begin{equation}
\begin{aligned}
    &|\Gamma^r - \Gamma^{r-1}| < \epsilon \quad \text{and} \\
    &\min\{U_T(\mathbf{m}^r),\, U_P(\mathbf{m}^r \mid \mathcal{D}_{obs})\} > \gamma
\end{aligned}
\end{equation}
where $\epsilon$ is the convergence tolerance and $\gamma$ is the minimum acceptable quality threshold. This dual condition ensures that convergence requires not only stability but also that neither objective is sacrificed below an acceptable level.

The first condition enforces Pareto stationarity: the solution has reached a point where further debate rounds yield diminishing marginal improvements to the joint objective. The second condition acts as a safety constraint, preventing degenerate equilibria where one utility collapses. Together, they define a bounded rational equilibrium, a stable state where neither agent can unilaterally improve their objective without degrading the other's, subject to the quality floor $\gamma$.

To prevent indefinite oscillation, we impose a maximum round budget $R_{max}$. If the convergence condition is not met within $R_{max}$ rounds, the Moderator selects the hypothesis with the highest $\Gamma^r$ across all rounds:
\begin{equation}
    \mathbf{m}^* = \underset{r \in \{1, \ldots, R_{max}\}}{\arg\max} \; \Gamma^r
\end{equation}

\subsection{Self-Validating Execution Strategy (SVE)}

To bridge the gap between refined model hypotheses and reliable code execution, we introduce SVE, a fully autonomous pipeline that structures code generation.

\subsubsection{Blueprint Construction}

The Architect module translates the equilibrium hypothesis $\mathbf{m}^*$ into a structured JSON blueprint $\mathcal{B}$, which serves as an intermediate representation between abstract modeling logic and concrete code. This design draws on the principle that separating specification from implementation reduces cascading errors in automated code synthesis~\cite{chen2023teaching}. Formally, $\mathcal{B}$ is defined as a tuple:
\begin{equation}
    \mathcal{B} = \left\langle \mathcal{V}, \mathcal{F}, \mathcal{I}, \mathcal{O} \right\rangle
\end{equation}
where $\mathcal{V} = \{(v_i, \tau_i, d_i)\}$ specifies variable declarations with types $\tau_i$ and tensor dimensions $d_i$; $\mathcal{F} = \{(f_j, \text{sig}_j, \text{dep}_j)\}$ defines function signatures with dependency relations; $\mathcal{I}$ specifies the data ingestion schema; and $\mathcal{O}$ defines the expected output format. This explicit intermediate representation decouples modeling decisions from syntactic implementation, reducing cascading errors during code generation.

\subsubsection{Automated Consistency Verification}

Rather than relying on costly human approval, we introduce a Verifier agent $\mathcal{A}_V$ that performs automated multi-level consistency checks between the blueprint $\mathcal{B}$ and the equilibrium hypothesis $\mathbf{m}^*$. The Verifier evaluates a set of formal consistency predicates and produces a composite verification score:
\begin{equation}
    V(\mathcal{B}, \mathbf{m}^*) = \prod_{p=1}^{P} v_p(\mathcal{B}, \mathbf{m}^*)
\end{equation}
where each $v_p \in \{0, 1\}$ is a binary consistency predicate. 

The verification outcome determines the downstream action:
\begin{equation}
    \mathbf{C}^* = \begin{cases} 
        \text{Builder}(\mathcal{B}), & \text{if } V(\mathcal{B}, \mathbf{m}^*) = 1 \\[6pt]
        \text{Builder}\!\left(\text{Revise}(\mathcal{B},\, \boldsymbol{\xi})\right), & \text{if } V(\mathcal{B}, \mathbf{m}^*) = 0
    \end{cases}
\end{equation}
where $\boldsymbol{\xi} = \{(p, \text{desc}_p) \mid v_p = 0\}$ is the set of violated predicates with diagnostic descriptions generated by $\mathcal{A}_V$. The Revise operator updates $\mathcal{B}$ by addressing each violation while preserving the components that passed verification. This process iterates for at most $T_{max}$ cycles:
\begin{equation}
    \mathcal{B}^{t+1} = \text{Revise}(\mathcal{B}^t, \boldsymbol{\xi}^t), \quad t = 0, 1, \ldots, T_{max} - 1
\end{equation}
terminating early when $V(\mathcal{B}^t, \mathbf{m}^*) = 1$. Compared to human gating, this automated mechanism achieves higher coverage on formal consistency dimensions that human reviewers frequently overlook, while eliminating latency and maintaining full autonomy.

\subsubsection{Self-Correcting Execution}

The synthesized code $\mathbf{C}^*$ executes within a sandboxed environment $\mathcal{E}$. In the event of a runtime failure, the error trace $\mathbf{e}^j$ is extracted and combined with the original code context to produce a corrected version:
\begin{equation}
    \mathbf{C}^{j+1} = \mathcal{F}_{refine}([\mathbf{C}^j \,;\, \mathbf{e}^j \,;\, \mathcal{B}])
\end{equation}
where $\mathcal{F}_{refine}$ denotes the LLM-based refinement operator and $j$ indexes the correction iteration. Crucially, we include the blueprint $\mathcal{B}$ in the refinement context to prevent the correction process from drifting away from the validated modeling intent. The refinement process terminates upon successful execution or after exhausting a maximum retry budget $J_{max}$.

We define the execution success indicator as:
\begin{equation}
    \mathds{1}_{exec} = \begin{cases} 1, & \text{if } \mathcal{E}(\mathbf{C}^j) \text{ returns valid output} \\ 0, & \text{otherwise} \end{cases}
\end{equation}

The overall Code Executability (CE) metric reported in our experiments is then computed as $\text{CE} = \frac{1}{|\mathcal{P}|}\sum_{p \in \mathcal{P}} \mathds{1}_{exec}^{(p)}$, where $\mathcal{P}$ is the problem set.

\subsection{Epistemic Self-Evolution}

To enable lifelong learning beyond a static knowledge base, we design a continuous self-evolution mechanism~\cite{voyager} that consolidates successful reasoning experiences into reusable knowledge.

\subsubsection{Knowledge Distillation}

Upon successfully solving a novel problem $\mathbf{q}_{new}$ with validated solution $\mathbf{s}_{new}^+$, a memory consolidation function $\Psi(\cdot)$ abstracts the solution into a generalized knowledge entry:
\begin{equation}
    k_{new} = \Psi(\mathbf{s}_{new}^+) = (\mathbf{x}_{new},\, \mathbf{c}_{new},\, \pi_{new})
\end{equation}
where $\mathbf{x}_{new}$ is the embedding of the problem, $\mathbf{c}_{new}$ is the validated code, and $\pi_{new}$ is the abstracted paradigm descriptor.

\subsubsection{Diversity-Aware Admission}

To prevent knowledge base bloating with redundant entries, we impose a diversity-aware admission criterion based on the maximum similarity with existing entries:
\begin{equation}
    \Delta(\mathbf{x}_{new}, \mathcal{K}) = \max_{\mathbf{x}_i \in \mathcal{K}} \frac{\mathbf{x}_{new}^\top \mathbf{x}_i}{\|\mathbf{x}_{new}\|_2 \cdot \|\mathbf{x}_i\|_2}
\end{equation}

The new entry is admitted if and only if it provides sufficient novelty:
\begin{equation}
    \mathcal{K}_{t+1} = \begin{cases}
        \mathcal{K}_t \cup \{k_{new}\}, & \text{if } \Delta(\mathbf{x}_{new}, \mathcal{K}_t) < \tau \\
        \mathcal{K}_t, & \text{otherwise}
    \end{cases}
\end{equation}

\begin{table*}[t]
\centering
{\footnotesize 
\begin{tabular}{ll|ccccc|c}
\toprule
\textbf{Method} & \textbf{Backbone} & \textbf{AE} & \textbf{MR} & \textbf{PS} & \textbf{RBA} & \textbf{CE (\%)} & \textbf{Avg.} \\ 
\midrule
\multirow{4}{*}{Direct Prompting} 
 & GPT-4o & 7.62$\pm$0.18 & 3.86$\pm$0.24 & 5.17$\pm$0.21 & 6.28$\pm$0.15 & 42.5 & 5.73 \\
 & Claude-3.5 & 7.78$\pm$0.16 & 4.15$\pm$0.22 & 5.42$\pm$0.19 & 6.50$\pm$0.17 & 45.8 & 5.96 \\
 & DeepSeek-R1 & 8.05$\pm$0.14 & 5.32$\pm$0.28 & 5.85$\pm$0.20 & 6.92$\pm$0.16 & 48.2 & 6.54 \\
 & Qwen-2.5-72B & 7.45$\pm$0.20 & 3.58$\pm$0.26 & 4.90$\pm$0.23 & 6.05$\pm$0.18 & 38.7 & 5.50 \\
\midrule
ReAct & GPT-4o & 7.89$\pm$0.15 & 5.24$\pm$0.31 & 6.02$\pm$0.19 & 6.71$\pm$0.22 & 51.8 & 6.47 \\
DS-Agent & GPT-4o & 8.18$\pm$0.14 & 7.08$\pm$0.26 & 7.47$\pm$0.18 & 7.86$\pm$0.20 & 68.2 & 7.65 \\
CAMEL & GPT-4o & 8.35$\pm$0.16 & 6.72$\pm$0.29 & 7.18$\pm$0.23 & 7.54$\pm$0.18 & 63.7 & 7.45 \\
AutoGen & GPT-4o & 8.42$\pm$0.13 & 6.95$\pm$0.25 & 7.62$\pm$0.17 & 7.93$\pm$0.19 & 71.4 & 7.73 \\
\midrule
\multirow{4}{*}{MM-Agent} 
 & GPT-4o & 9.15$\pm$0.11 & 7.28$\pm$0.22 & 8.44$\pm$0.14 & 8.85$\pm$0.12 & 82.0 & 8.43 \\
 & Claude-3.5 & 9.22$\pm$0.10 & 7.45$\pm$0.20 & 8.52$\pm$0.13 & 8.90$\pm$0.11 & 84.5 & 8.52 \\
 & DeepSeek-R1 & 9.30$\pm$0.09 & 7.82$\pm$0.18 & 8.68$\pm$0.12 & 9.02$\pm$0.10 & 85.2 & 8.71 \\
 & Qwen-2.5-72B & 8.95$\pm$0.13 & 6.90$\pm$0.24 & 8.15$\pm$0.16 & 8.58$\pm$0.14 & 76.8 & 8.15 \\
\midrule
Human Expert & -- & 9.04$\pm$0.35 & 7.91$\pm$0.42 & 7.42$\pm$0.38 & 8.92$\pm$0.30 & -- & 8.32 \\
\midrule
\multirow{4}{*}{\textbf{Sci-Mind (Ours)}} 
 & GPT-4o & \cellcolor{third}9.45$\pm$0.09 & \cellcolor{third}8.92$\pm$0.14 & \cellcolor{third}9.15$\pm$0.11 & \cellcolor{third}9.05$\pm$0.10 & \cellcolor{third}96.4 & \cellcolor{third}9.14 \\
 & Claude-3.5 & \cellcolor{second}9.50$\pm$0.08 & \cellcolor{second}9.05$\pm$0.12 & \cellcolor{second}9.22$\pm$0.10 & \cellcolor{second}9.12$\pm$0.09 & \cellcolor{second}97.1 & \cellcolor{second}9.22 \\
 & DeepSeek-R1 & \cellcolor{first}\textbf{9.58}$\pm$0.07 & \cellcolor{first}\textbf{9.28}$\pm$0.10 & \cellcolor{first}\textbf{9.35}$\pm$0.09 & \cellcolor{first}\textbf{9.25}$\pm$0.08 & \cellcolor{first}\textbf{97.8} & \cellcolor{first}\textbf{9.37} \\
 & Qwen-2.5-72B & 9.28$\pm$0.11 & 8.52$\pm$0.16 & 8.82$\pm$0.13 & 8.78$\pm$0.12 & 93.5 & 8.85 \\
\bottomrule
\end{tabular}%
}
\caption{\label{tab:main}
Main results on MM-Bench across four backbones. All scores are reported as mean $\pm$ std over three evaluation runs.}
\end{table*}

\section{Experiments}
 
\subsection{Experimental Setup}

\textbf{Datasets.}
We evaluate on two complementary benchmarks. MM-Bench~\cite{mm_agent} contains 111 real-world mathematical modeling problems from MCM/ICM competitions spanning 2000 to 2025, covering 10 domains and 8 task types with end-to-end problem solving requirements. EngiBench Level 3~\cite{engibench} provides 43 open-ended engineering modeling tasks from MCM/ICM, CUMCM, and APMCM spanning 2010 to 2024, covering Systems \& Control, Physical \& Structural, and Chemical \& Biological subfields with official rubrics.

\textbf{Baselines.}
We compare against seven baselines spanning direct prompting, specialized agents, and multi-agent frameworks. Direct Prompting applies Zero-shot CoT~\cite{cot} to the backbone LLM. ReAct~\cite{react} interleaves reasoning with acting. DS-Agent~\cite{ds_agent} leverages case-based reasoning from Kaggle. CAMEL~\cite{camel} enables role-playing multi-agent communication. AutoGen~\cite{autogen} supports flexible multi-agent conversation with code execution. MM-Agent~\cite{mm_agent} is the current state-of-the-art framework for mathematical modeling. Human Expert solutions are those awarded Honorable Mention or higher in actual MCM/ICM competitions.

\textbf{Backbones.}
We evaluate Sci-Mind and key baselines across four representative LLMs: GPT-4o~\cite{gpt4}, Claude-3.5-Sonnet~\cite{claude}, DeepSeek-R1~\cite{deepseek_r1}, and Qwen-2.5-72B~\cite{qwen}. 

\textbf{Metrics.}
Following the MM-Bench evaluation protocol~\cite{mm_agent}, we report four qualitative dimensions scored on a 1 to 10 scale: Analysis Evaluation (AE), Modeling Rigorousness (MR), Practicality (PS), and Result Analysis (RBA). The Avg. score is their arithmetic mean. We additionally report Code Executability (CE), defined as the percentage of generated solutions that execute without runtime errors. Each problem is evaluated 3 times and we report the mean with standard deviation. For EngiBench, we follow its official rubric-based protocol~\cite{engibench}.

\begin{table*}[t]
\centering
{\footnotesize 
\begin{tabular}{ll cccccc c}
\toprule
\multirow{2}{*}{\textbf{Method}} & \multirow{2}{*}{\textbf{Backbone}} & \multicolumn{2}{c}{\textbf{SC}} & \multicolumn{2}{c}{\textbf{PS}} & \multicolumn{2}{c}{\textbf{CB}} & \multirow{2}{*}{\textbf{Avg.}} \\
\cmidrule(lr){3-4} \cmidrule(lr){5-6} \cmidrule(lr){7-8} 
 & & \textbf{RS} & \textbf{CE} & \textbf{RS} & \textbf{CE} & \textbf{RS} & \textbf{CE} & \\
\midrule
\multirow{2}{*}{Direct Prompting}
 & GPT-4o & 28.4$\pm$1.8 & 35.0 & 24.1$\pm$2.1 & 30.0 & 31.7$\pm$1.6 & 40.0 & 28.1 \\
 & DeepSeek-R1 & 34.2$\pm$1.5 & 42.5 & 30.8$\pm$1.9 & 37.5 & 38.5$\pm$1.4 & 47.5 & 34.5 \\
\midrule
\multirow{2}{*}{MM-Agent}
 & GPT-4o & 51.3$\pm$1.2 & 72.5 & 46.8$\pm$1.5 & 65.0 & 54.0$\pm$1.1 & 75.0 & 50.7 \\
 & DeepSeek-R1 & \cellcolor{third}56.8$\pm$1.0 & \cellcolor{third}77.5 & \cellcolor{third}52.4$\pm$1.3 & \cellcolor{third}72.5 & \cellcolor{third}59.2$\pm$0.9 & \cellcolor{third}80.0 & \cellcolor{third}56.1 \\
\midrule
\multirow{2}{*}{Sci-Mind}
 & GPT-4o & \cellcolor{second}63.8$\pm$0.8 & \cellcolor{second}90.0 & \cellcolor{second}58.5$\pm$1.0 & \cellcolor{second}85.0 & \cellcolor{second}66.2$\pm$0.7 & \cellcolor{second}92.5 & \cellcolor{second}62.8 \\
 & DeepSeek-R1 & \cellcolor{first}\textbf{68.5}$\pm$0.6 & \cellcolor{first}\textbf{92.5} & \cellcolor{first}\textbf{64.2}$\pm$0.8 & \cellcolor{first}\textbf{90.0} & \cellcolor{first}\textbf{71.0}$\pm$0.5 & \cellcolor{first}\textbf{95.0} & \cellcolor{first}\textbf{67.9} \\
\bottomrule
\end{tabular}%
}
\caption{\label{tab:engibench}
Evaluation on EngiBench Level 3. SC: Systems \& Control, PS: Physical \& Structural, CB: Chemical \& Biological. RS denotes rubric score (\%), CE denotes code executability (\%). RS is reported as mean $\pm$ std over three runs.
}
\end{table*}
 
\subsection{Main Results}

Table~\ref{tab:main} presents the main results on MM-Bench. Three key findings emerge.

\textbf{Consistent SOTA across backbones.} Sci-Mind outperforms all baselines on every backbone. With GPT-4o it surpasses MM-Agent by 0.71 in Avg. score. With DeepSeek-R1 the margin remains at 0.66. Sci-Mind with Qwen-2.5-72B achieves 8.85, still exceeding MM-Agent with DeepSeek-R1 at 8.71, demonstrating that the gains are architecture-driven rather than backbone-dependent.

\textbf{MR and CE as primary improvement channels.} Across all backbones, the largest gains appear in MR with an average improvement of 1.52 and CE with an average improvement of 13.8, confirming that ACD and EMR address the two core bottlenecks, theoretical drift and the abstraction-implementation gap, that backbone scaling alone cannot resolve. DeepSeek-R1 amplifies these advantages further, reaching MR of 9.28 and CE of 97.8.

\textbf{Generic multi-agent frameworks are insufficient.} CAMEL and AutoGen outperform single-agent baselines but fall short of the specialized MM-Agent, indicating that general-purpose multi-agent dialogue lacks the targeted verification needed for scientific modeling.

Regarding Human Experts, Sci-Mind's advantage is most pronounced in PS, scoring 9.15 compared to 7.42, likely because the system validates implementation details against data constraints more exhaustively. We note that all scores are produced by automated evaluators and may not fully capture the nuanced creativity of human solutions.

\subsection{Cross-Domain Generalization}
 
Table~\ref{tab:engibench} confirms that Sci-Mind's advantages generalize to engineering domains beyond competition-style problems. With GPT-4o, Sci-Mind achieves an average RS improvement of 12.1 over MM-Agent across three subfields. With DeepSeek-R1, the gap remains at 11.8. The largest gains appear in PS, where complex dynamical modeling and spatially heterogeneous data handling are required, precisely the scenarios where ACD most effectively identifies model-data mismatches. The smaller gap in CB reflects the heavier reliance on well-established reaction kinetics paradigms that baseline agents already handle reasonably well. CE improvements average 17.5 over MM-Agent across all configurations, indicating that EMR's structural priors transfer robustly across domain boundaries. Notably, Sci-Mind with GPT-4o already surpasses MM-Agent with DeepSeek-R1 on both RS and CE in every subfield, further confirming that the framework's design contributes more than backbone strength alone.

\begin{table*}[t]
\centering
{\footnotesize 
\begin{tabular}{c|ccccc|c|c}
\toprule
\textbf{Variant} & \textbf{AE} & \textbf{MR} & \textbf{PS} & \textbf{RBA} & \textbf{CE (\%)} & $\Delta$\textbf{CE} & \textbf{Overall} \\
\midrule
\multicolumn{8}{c}{\cellcolor[HTML]{EEEEEE}\textit{Module-Level Ablation}} \\
w/o EMR & 9.20$\pm$0.15 & 8.50$\pm$0.18 & 7.80$\pm$0.22 & 8.10$\pm$0.19 & 55.2 & $-$41.2 & 8.40 \\
w/o ACD & 9.30$\pm$0.12 & 7.45$\pm$0.24 & 8.60$\pm$0.16 & 8.50$\pm$0.15 & 94.0 & $-$2.4 & 8.46 \\
w/o SVE & \cellcolor{second}9.38$\pm$0.10 & \cellcolor{second}8.78$\pm$0.15 & 8.45$\pm$0.17 & 8.62$\pm$0.14 & 79.5 & $-$16.9 & \cellcolor{second}8.81 \\
\midrule
\multicolumn{8}{c}{\cellcolor[HTML]{EEEEEE}\textit{ACD Design Alternatives}} \\
Self-Reflection & 9.32$\pm$0.11 & 7.68$\pm$0.21 & 8.72$\pm$0.14 & 8.55$\pm$0.13 & 93.5 & $-$2.9 & 8.57 \\
Homogeneous & 9.28$\pm$0.12 & 7.92$\pm$0.19 & 8.65$\pm$0.15 & 8.60$\pm$0.14 & \cellcolor{third}94.2 & \cellcolor{third}$-$2.2 & 8.61 \\
w/o Moderator & \cellcolor{third}9.35$\pm$0.10 & 8.35$\pm$0.16 & \cellcolor{third}8.82$\pm$0.13 & \cellcolor{second}8.72$\pm$0.12 & \cellcolor{second}95.0 & \cellcolor{second}$-$1.4 & \cellcolor{second}8.81 \\
\midrule
\multicolumn{8}{c}{\cellcolor[HTML]{EEEEEE}\textit{EMR Retrieval Modality}} \\
Text-Only & 9.25$\pm$0.13 & \cellcolor{third}8.55$\pm$0.17 & 8.12$\pm$0.19 & 8.35$\pm$0.16 & 65.8 & $-$30.6 & 8.57 \\
Code-Only & 9.18$\pm$0.14 & 7.90$\pm$0.20 & \cellcolor{second}8.85$\pm$0.12 & \cellcolor{third}8.65$\pm$0.13 & 91.5 & $-$4.9 & \cellcolor{third}8.65 \\
\midrule
\textbf{Sci-Mind (Ours)} & \cellcolor{first}\textbf{9.45}$\pm$0.09 & \cellcolor{first}\textbf{8.92}$\pm$0.14 & \cellcolor{first}\textbf{9.15}$\pm$0.11 & \cellcolor{first}\textbf{9.05}$\pm$0.10 & \cellcolor{first}\textbf{96.4} & \cellcolor{first}\textemdash & \cellcolor{first}\textbf{9.14} \\
\bottomrule
\end{tabular}%
}
\caption{\label{tab:ablation}
Comprehensive ablation on MM-Bench with GPT-4o backbone. All scores are reported as mean $\pm$ std over three evaluation runs.
}
\end{table*}
 
\subsection{Convergence Behavior of ACD}

\begin{figure}
    \centering
    \includegraphics[width=1\linewidth]{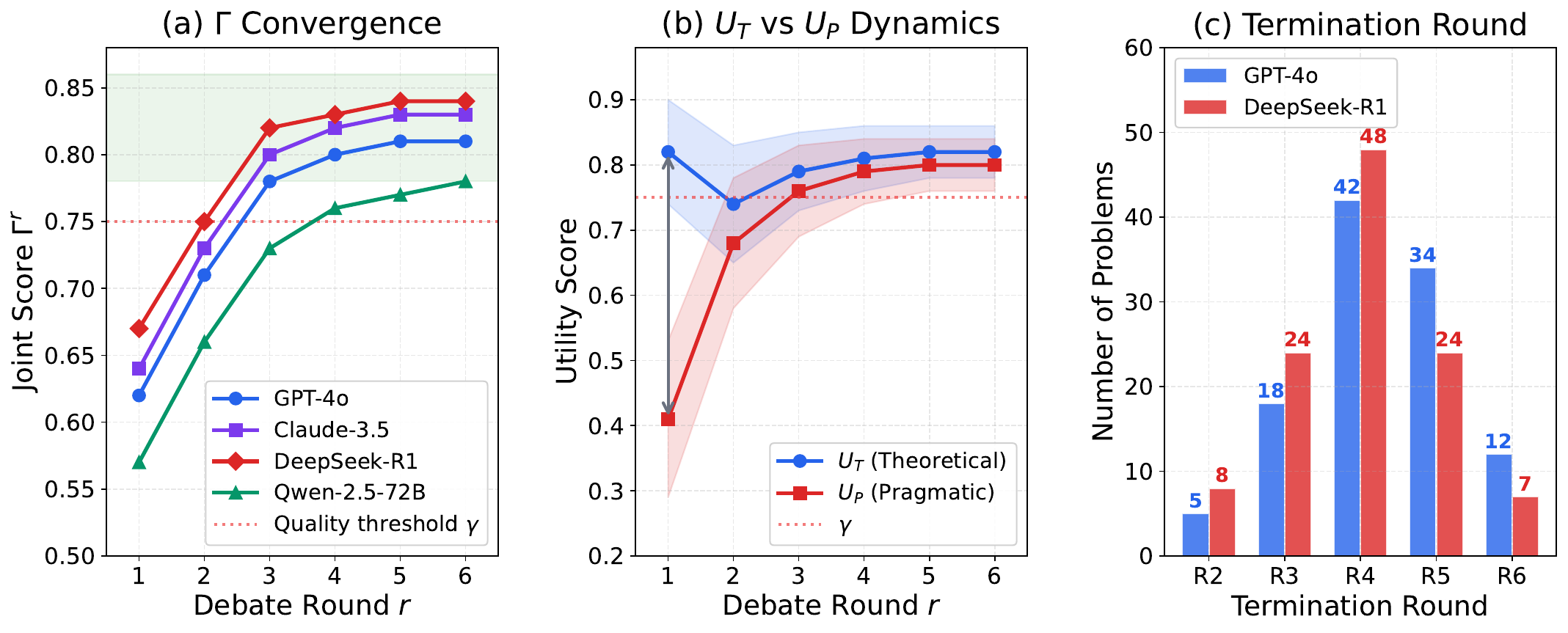}
\caption{ACD convergence analysis across 111 MM-Bench problems.} \label{fig3}
\end{figure}

Figure~\ref{fig3} presents the ACD convergence dynamics across all 111 MM-Bench problems. All four backbones exhibit the same qualitative trajectory in panel (a): $\Gamma$ rises steeply from Round 1 to Round 3 then plateaus into the convergence zone, with DeepSeek-R1 converging fastest at 3.6 rounds on average with 94\% convergence rate and Qwen-72B requiring 4.8 rounds at 82\%. This ordering aligns with backbone reasoning strength, yet even DeepSeek-R1 starts with $U_P$ well below the quality threshold $\gamma$, confirming that adversarial verification remains essential regardless of backbone capability. Panel (b) reveals the underlying mechanism under GPT-4o: at Round 1, $U_T$ reaches 0.82 while $U_P$ is only 0.41, creating a large theory-data gap where the Theorist generates mathematically elegant but practically infeasible models. The Pragmatist's critique rapidly closes this gap in Round 2 by forcing $U_P$ up to 0.68 at the cost of a temporary $U_T$ dip to 0.74 as the Theorist simplifies its formulation to accommodate data constraints. From Round 3 onward both utilities co-ascend with shrinking variance, indicating that later rounds primarily resolve edge cases rather than systematic modeling errors. Panel (c) confirms that the modal termination round is R4 for both GPT-4o and DeepSeek-R1, with only 12 and 7 problems respectively requiring the full $R_{max} = 6$ budget, suggesting that the computational overhead of multi-round debate is bounded and predictable in practice.
 
\subsection{Ablation Study}

Table~\ref{tab:ablation} presents the ablation results across three tiers.

\textbf{RQ1: Does EMR bridge the abstraction-implementation gap?}
Removing EMR causes CE to drop from 96.4\% to 55.2\% while MR decreases only by 0.42. This asymmetry confirms that ACD preserves modeling knowledge but EMR alone supplies the procedural scaffolding for translating equations into executable code. The retrieval modality ablation further disentangles each knowledge tier: Text-Only retrieval maintains MR at 8.55 but CE falls to 65.8\% because the agent lacks implementation templates, while Code-Only retrieval achieves CE of 91.5\% but MR drops to 7.90 because conceptual guidance for model selection is absent. These complementary degradation patterns validate that the tri-level representation is essential and each tier addresses a distinct failure mode.

\textbf{RQ2: Does ACD escape theoretical local optima?}
Removing ACD yields the sharpest MR decline of 1.47 while CE stays at 94.0\%, confirming that without adversarial verification the agent produces executable but theoretically flawed models. The design alternatives form a monotonic progression: Self-Reflection reaches MR 7.68, Homogeneous Debate 7.92, ACD without Moderator 8.35, and full ACD 8.92. The gap of 1.00 between Homogeneous Debate and full ACD confirms that objective asymmetry is the critical factor, and the Moderator contributes an additional 0.57 by preventing premature termination on suboptimal hypotheses.

\textbf{RQ3: Does SVE enhance execution robustness?}
Removing SVE drops CE by 16.9\% with only 0.14 MR change, confirming its dedicated role as a verification layer downstream of modeling decisions. This pattern complements the ACD ablation: ACD ensures theoretical soundness while SVE ensures faithful translation into executable code through formal consistency checking.

\subsection{Qualitative Case Study}

\begin{figure*}
    \centering
    \includegraphics[width=1\linewidth]{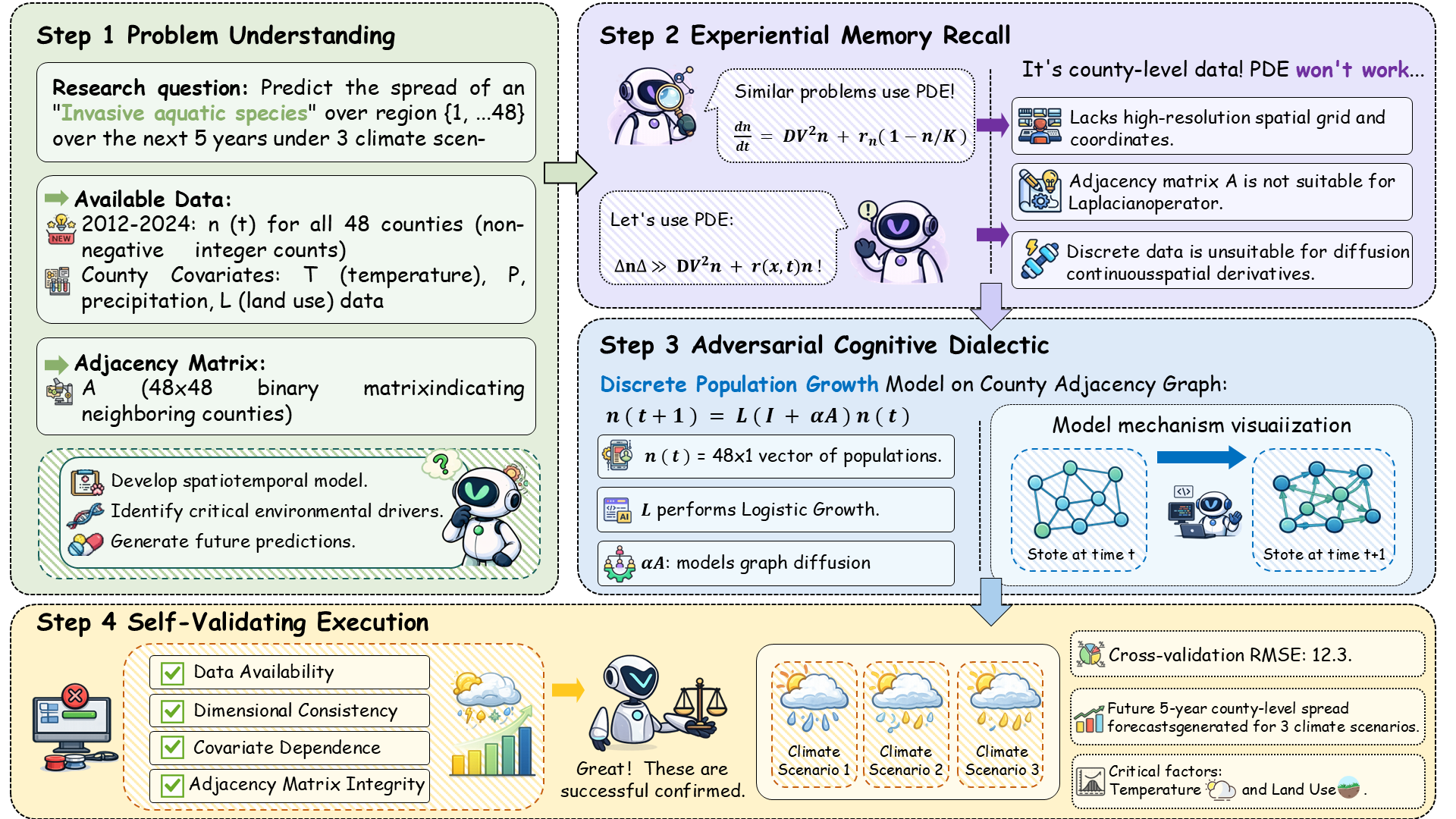}
\caption{Case study reasoning trace of Sci-Mind.} \label{fig4}
\end{figure*}

Figure~\ref{fig4} traces how Sci-Mind resolves a
representative modeling challenge, where
the task is to predict invasive species spread across 48
counties under three climate scenarios. The dataset provides
county-level population counts, environmental covariates,
and a binary adjacency matrix, but no continuous spatial
coordinates. EMR retrieves historically similar cases that
employed a Fisher-KPP PDE, which would be the natural choice
for spatial diffusion modeling. However, the system
simultaneously recognizes a fundamental incompatibility: the
dataset lacks the high-resolution spatial grid required by
the Laplacian operator, the adjacency matrix encodes only
binary contiguity rather than metric distances, and discrete
county-level observations cannot support continuous spatial
derivatives. Rather than propagating this infeasible
formulation into code, the Adversarial Cognitive Dialectic
drives the Theorist to reformulate the dynamics as a discrete
population growth model on the county adjacency graph:
$\mathbf{n}(t\!+\!1) = \mathbf{L}(\mathbf{I}+\alpha
\mathbf{A})\mathbf{n}(t)$, where the adjacency matrix serves
directly as the dispersal operator and environmental
covariates modulate local logistic growth rates. The
Self-Validating Execution module then confirms data
availability, dimensional consistency, covariate dependence,
and adjacency matrix integrity before code generation. The
model executes successfully, generating 5-year county-level
spread forecasts for three climate scenarios with
cross-validation RMSE of 12.3.

\section{Discussion and Conclusion}

We presented Sci-Mind, a cognitively-inspired framework that integrates Experiential Memory Recall, Adversarial Cognitive Dialectic, and Self-Validating Execution Strategy for autonomous mathematical modeling. Experiments on MM-Bench and EngiBench across four backbones confirm state-of-the-art performance in both modeling rigorousness and code executability, with gains orthogonal to backbone capabilities.

Our analysis reveals two broader insights. First, objective asymmetry matters more than agent count for effective verification: the gap between Homogeneous Debate and full ACD far exceeds that between Self-Reflection and Homogeneous Debate, suggesting that decomposing verification into competing objectives is more effective than scaling homogeneous reviewers. Second, the abstraction-implementation gap is a distinct failure mode where agents can formulate correct models but cannot implement them, which retrieval of executable structural priors directly addresses.

One limitation is that our experiments focus on competition-style modeling tasks, and generalizability to industrial or laboratory workflows remains to be validated. This motivates two future directions: extending the fixed verification predicates to a learnable mechanism that adapts to new domains, and accumulating domain-specific critique templates through finer-grained self-evolution to accelerate convergence on specialized tasks. We hope Sci-Mind provides a useful step toward autonomous agents that reason about scientific problems with the rigor and groundedness of human domain experts.

\bibliography{tacl2021}
\bibliographystyle{acl_natbib}

\iftaclpubformat

\onecolumn






  

\end{document}